\documentclass{PoS}

\title{Lattice Field Theory Methods in Modern Biophysics}

\ShortTitle{Lattice Field Theory Methods in Modern Biophysics}

\author{\speaker{Anthony Duncan}\\
        Department of Physics and Astronomy, University of Pittsburgh, Pittsburgh, PA 15260\\
        E-mail: \email{tony@dectony.phyast.pitt.edu}}

\abstract{An effective field theory exists describing a very large class of biophysically
interesting Coulomb gas systems: the lowest order (mean-field) version of
this theory takes the form of a generalized Poisson-Boltzmann theory.
Interaction terms depend on details (finite-size effects, multipole properties, etc).
Convergence of the loop expansion holds only if mutual interactions of mobile
charges are small compared to their interaction with the fixed-charge
environment, which is frequently not the case. Problems with the strongly-coupled
effective theory can be circumvented with an alternative local lattice formulation,
with real positive action. In realistic situations, with variable dielectric, a determinant
of the Poisson operator must be inserted to generate correct electrostatics. Methods
adopted from unquenched lattice QCD do this very efficiently.}

\FullConference{XXIVth International Symposium on Lattice Field Theory\\
                July 23-28, 2006\\
                Tucson, Arizona, USA}

\input psfig.sty
\begin{document}

\section{Introduction}

  A central paradigmatic problem of modern biophysics centers on the thermodynamic
  properties of {\em generalized Coulomb gases}, by which is meant a system of
  both mobile and fixed electric charges (or more generally, particles with higher
  multipole moments as well), in a background (typically water, but also 
  electrically neutral interiors of protein polypeptides, in which fixed charges may
  be embedded) of  variable dielectric. Direct molecular dynamics simulations of
  Coulomb gas problems are impractical for large systems, as the long-range
  character of the Coulomb interaction means that the electrostatic energy of every
  pair of particles in the system has to be computed. In systems with uniform 
  dielectric, this problem can be ameliorated by Fourier (Ewald) techniques, but
  in the general case, the dielectric ``constant" in fact varies spatially (and also dynamically,
  in the course of the simulation, if macroions or polymer components are allowed to
  change conformation), making Fourier methods impractical.
  
  Of course the screening effects in a
  homogeneous medium with mobile charges have been understood for a long time
  on the basis of Debye-H\"uckel theory \cite{DebHuck}, provided the charge concentrations are
  not too high. Here one starts from a Poisson-Boltzmann equation which summarizes the
  mean-field effects of the mobile ions on each other (and with any fixed charges).
  Starting in the 1940s,
  a great deal was accomplished with a linearized version of this equation (the DLVO
  method introduced by Derjaguin, Landau, Verwey and Overbeek \cite{DLVO}). However,
  such methods do not provide any intrinsic procedure for systematic improvement of the
  mean-field result, and frequently fail completely in the regime of high concentrations.
  
    A more general formalism for dealing with Coulomb gases under very general
    circumstances was introduced in the early 90's \cite{CoalDunc}: the initial 
    emphasis was in dealing with systems of fixed charged macroions surrounded by
    a gas of small mobile ions. The grand canonical partition function for such systems
    can be converted into a path-integral regularized on a spatial lattice, and the
    saddle-point expansion of the functional integral then leads to a (discretized)
    Poisson-Boltzmann equation which can be rapidly solved numerically.  Moreover,
    the basic technique allows for straightforward generalization to systems with
    short-range repulsive forces between the mobile ions \cite{finsize}, multipolar ions \cite{multipol}, 
    charged polymer interactions \cite{electrolyte,partitioning1,partitioning2}, among others.
    In all of these cases the higher order fluctuation (``loop") effects are clearly defined,
    and the leading corrections to the mean-field result computable.
    Despite the formally attractive nature of this effective field approach, there still remains
    the difficulty that in many cases these fluctuation corrections are very large, so a
    perturbative saddle-point expansion does not yield useful results. Furthermore, the
    effective action for these theories is always complex, so direct Monte Carlo simulation of 
    the functional integral is impractical due to the infamous sign effect.
    
      Recently, Maggs and collaborators suggested \cite{Maggs1} an alternative approach
      to Coulomb gases in which the long-range Coulomb interaction is localized by
      writing the path integral for the partition function in terms of local electric field variables
      (essentially, one takes the Hamiltonian
      path integral for  finite temperature Maxwell electrodynamics and neglects magnetic terms).
      A considerable amount of work has now been devoted to streamlining and improving the
      efficiency of this approach \cite{CoalDunSedge1,fftpaper}. The original algorithm of
      Maggs et al. only handles systems of uniform dielectric, however, for which one may argue
      that Fourier accelerated molecular dynamics simulations are competitive. For systems in
      which the dielectric medium is dynamical, the Maggs et al. functional integral produces
      a spurious interaction force between the particles, which must be removed to 
      obtain the correct electrostatic energy. Recent work \cite{multibos1,multibos2}
      has shown that this can be done
      exactly by introducing the determinant of the generalized Poisson operator into
      the path integral, in complete analogy to the way the determinant of the quark Dirac
      operator must be introduced in the path integral of unquenched QCD. Numerical
      simulations can be performed using a variety of methods imported from lattice QCD:
      in the following, we describe results for the structure factor of a dielectric plasma
      obtained using the L\"uscher multiboson method.

\section{ Functional Formalism for Coulomb Gases }     

Although the Coulomb interaction is long-range, it has the special,
and extremely useful, feature that it is the Green's function for
a local operator, the Laplacian. This makes it possible to
remove the  nonlocal Coulomb interaction by introducing a local auxiliary field-
\begin{eqnarray*}
   &&   e^{-\frac{\beta}{2\epsilon}\int d\vec{r}d\vec{r}^{\prime}\rho(\vec{r})\rho(\vec{r}^{\prime})/|\vec{r}-\vec{r}^{\prime}|}\\
&=& C\int {\cal D}\chi e^{\frac{\epsilon}{8\pi\beta}\int\chi\Delta\chi d\vec{r}+i\int\chi(\vec{r})\rho(\vec{r})d\vec{r}}
\end{eqnarray*}

Here $\rho(\vec{r})$ is the charge density: to be specific, imagine a system with stationary charges 
 (charge $Q_{j}e$, locations $\vec{R}_{j}$) and mobile simple ions (charges $\pm e$, locations $\vec{r}_{k},\vec{r}_{l}$)
 so the charge density is

\begin{eqnarray*}
  \rho(\vec{r}) &=& -e\sum_{j}Q_{j}\delta(\vec{r}-\vec{R}_{j})+e\sum_{k}\delta(\vec{r}-\vec{r}_{k}) \\
  &-&e\sum_{l}\delta(\vec{r}-\vec{r}_{l})
\end{eqnarray*}

The full Hamiltonian in a typical Coulomb gas consists of a Coulomb energy term, plus single-particle exclusion potentials $V(\vec{r}_{k,l})$;
\begin{eqnarray*}
  H&=&\frac{1}{2\epsilon}\int d\vec{r}d\vec{r}^{\prime}\frac{\rho(\vec{r})\rho(\vec{r}^{\prime})}{|\vec{r}
  -\vec{r}^{\prime}|}  \\
  &+&\frac{1}{\beta}\sum_{k}V(\vec{r}_{k})+\frac{1}{\beta}\sum_{l}V(\vec{r}_{l})
  \end{eqnarray*}
  
It is most convenient to work in the grand-canonical framework, so we
introduce chemical potentials $\mu_{\pm 1}$, and the partition function becomes
\begin{eqnarray*}
  Z_{\rm gc}=\sum_{n_{k},n_{l}}\frac{1}{n_{k}!n_{l}!}\int\prod d\vec{r}_{k}d\vec{r}_{l}e^{-\beta H}
  \end{eqnarray*}
  
 In addition to localizing the Coulomb potential, the auxiliary field representation
 factorizes the partition sums:
 \begin{eqnarray*}
  \sum_{n_{k},n_{l}}\int\frac{d\vec{r}_{k}d\vec{r}_{l}}{n_{k}!n_{l}!}\exp{(\sum_{k}(ie\chi(\vec{r}_{k})-V(\vec{r}_{k})))}  \\
  \cdot\exp{(-\sum_{l}(ie\chi(\vec{r}_{l})+V(\vec{r}_{l}))+\beta(\mu_{+}n_{k}+\mu_{-}n_{l}))}\\
  =\exp{(e^{\beta\mu_{+}}\int d\vec{r}e^{(ie\chi-V)}+e^{\beta\mu_{-}}\int d\vec{r}e^{(-ie\chi-V)})}
  \end{eqnarray*}

and we obtain an exact path integral representation for $Z_{\rm gc}$:
\begin{eqnarray}
&&Z_{\rm gc}=\int{\cal D}\chi(\vec{r})\exp{(\frac{\alpha}{2}\int\chi(\vec{r})\Delta\chi(\vec{r}) d\vec{r})} \nonumber \\
&&\cdot\exp{ (\int(\gamma_{+}e^{i\chi-V}+\gamma_{-}e^{-i\chi-V})d\vec{r})} \nonumber \\
&&\cdot\exp{(-i\sum_{j}Q_{j}\chi(\vec{R}_{j}))} \nonumber \\
\label{pathintegralzgc}
&\equiv& \int{\cal D}\chi(\vec{r})\exp{-S(\chi)}
 \end{eqnarray}
 where $\alpha\equiv\frac{\epsilon}{4\pi\beta e^{2}},\gamma_{\pm}\equiv e^{\beta\mu_{\pm}}$.
 Although we have written the path integral in terms of fields defined for continuous space,
 one needs to regulate the preceding (and following) formulas on a spatial lattice to have
 a well-defined theory.
  
  \subsection{Mean-field Theory and Loop Expansion}
  
 The grand-canonical partition function $Z_{\rm gc}=\int {\cal D}\chi\exp{-S(\chi)}$ has a complex saddle-point
 of the Hubbard-Stratonovich type:
 \begin{eqnarray*}
    \chi &=&  -i\bar{\phi}+\xi \\
    \frac{\delta S}{\delta \chi(\vec{r})}(\chi=-i\bar{\phi}) &=& 0
    \end{eqnarray*}
   The saddle-point condition is the Poisson-Boltzmann equation for the system:
    \begin{eqnarray*}
    \alpha\Delta\bar{\phi}(\vec{r})&=&\gamma_{+}e^{\bar{\phi}(\vec{r})-V(\vec{r})}
    +\gamma_{-}e^{-\bar{\phi}(\vec{r})-V(\vec{r})}  \\
    &-&\sum_{j}Q_{j}\delta(\vec{r}-\vec{R}_{j})
    \end{eqnarray*}
 In general, this is solved rapidly and accurately by relaxation techniques. Evaluating the 
 action at the saddle-point gives the
  Thermodynamic Potential $\Xi\equiv\log{Z_{\rm gc}}$ in leading order:
  \begin{eqnarray}
  \label{meanXi}
    \Xi_{\rm mf}&=&-\frac{\alpha}{2}\int \bar{\phi}\Delta\bar{\phi}d\vec{r} \nonumber  \\
    &+&\gamma_{+}\int e^{\bar{\phi}-V}d\vec{r}+\gamma_{-}\int e^{-\bar{\phi}-V} \nonumber \\
    &-&\sum_{j}Q_{j}\bar{\phi}(\vec{R}_{j})
    \end{eqnarray}
 In practice, all spatial integrals and differential operators are regulated on a spatial lattice (as
 mentioned previously). 
\subsection{Systematic Loop Expansion of Thermodynamic Potential $\Xi$}

The mean-field approximation (\ref{meanXi}) is only useful if the
fluctuation corrections are small: in fact, if they are, we have a systematic procedure
for improving the mean-field result. Expanding around the saddle-point (mean-field) solution, $\chi = -i\bar{\phi}+\xi$:

\begin{eqnarray*}
\Xi &=& \Xi_{\rm mf}+\Xi_{\rm loops} 
\end{eqnarray*}
where 
\begin{eqnarray*}
\Xi_{\rm loops} &=& \log{(\int{\cal D}\xi \exp{-(S_{0}(\xi)+S_{\rm int}(\xi))})}
\end{eqnarray*}
The mean-field action is 
\begin{eqnarray*}
S_{0}(\xi) &=& -\frac{\alpha}{2}\int \xi\Delta\xi d\vec{r}   \\
&+&\frac{1}{2}\int(\gamma_{+}e^{(\bar{\phi}-V)}+\gamma_{-}e^{(-\bar{\phi}-V)})
d\vec{r} 
\end{eqnarray*}
and the fluctuation terms (through order $\xi^{4}$) are
\begin{eqnarray*}
S_{\rm int}(\xi)&=&\frac{-i}{3!}\int(\gamma_{+}e^{(\bar{\phi}-V)}-\gamma_{-}e^{(-\bar{\phi}-V)})\xi^{3}d\vec{r} \\
 &+&\frac{1}{4!}\int (\gamma_{+}e^{(\bar{\phi}-V)}+\gamma_{-}e^{(-\bar{\phi}-V)})\xi^{4}d\vec{r}
\end{eqnarray*} 
The loop (i.e. fluctuation) corrections involve a propagator defined as
\begin{eqnarray*}
  D(\vec{r},\vec{r}^{\prime})\equiv <\vec{r}|(-\alpha\Delta+\gamma_{+}e^{\bar{\phi}-V}+\gamma_{-}e^{-\bar{\phi}-V})^{-1}|\vec{r}^{\prime}>
  \end{eqnarray*}
  Unlike the situation in conventional field-theoretical perturbative expansions, in typical Coulomb
  gas applications this propagator cannot be written down analytically, as the system involves
  fixed charges in essentially arbitrary locations (so Fourier methods fail). Instead, the
  propagator is computed (on a lattice) numerically. Once this is done,
   1-loop, 2-loop, etc. corrections to the thermodynamic potential follow immediately:
  \begin{eqnarray*}
  \Xi^{(1)}&=& -\frac{1}{2}\log{{\rm det} D} \\
  \Xi^{(2)}&=& \frac{1}{8}\int (\gamma_{+}e^{(\bar{\phi}-V)}+\gamma_{-}e^{(-\bar{\phi}-V)})D^{2}(\vec{r},\vec{r})d\vec{r}
  \end{eqnarray*}
  
  The formalism described above has been applied to numerous problems, including the 
  thermodynamics of charged polymers in electrolytes \cite{electrolyte} and the partitioning
  of charged polymers between cavities \cite{partitioning1,partitioning2}.
\subsection{Extensions of the Formalism}

The path integral formalism for Coulomb  gases outlined above is extremely flexible:
one may easily modify it to include
\begin{enumerate}
\item Non-Coulomb pairwise interactions (such as repulsive Yukawa) of the mobile charges
\cite{finsize}.
\item Higher multipoles (dipole, quadrupole, etc) on the mobile charges \cite{multipol}.
\end{enumerate}

 In order to introduce a repulsive short-range core interaction between the mobile
 charges (in addition to the long range Coulomb piece) we consider \cite{finsize} a Coulomb/Yukawa Gas,
 with the interaction energy of mobile charges given by:
 \newpage
 
\begin{eqnarray*}
V_{\rm pair} &=& \frac {1}{2 \epsilon} \int d\vec{r} \, d\vec{r'} \rho (\vec{r})
\frac {1}{|\vec{r} - \vec{r'} |} \rho (\vec{r'})   \\
& +&
\frac {C k_B T \lambda_B}{2} \int d\vec{r} \, d\vec{r'} \rho_s (\vec{r}) 
\frac {e^{- \kappa | \vec{r} - \vec{r'}|}}{| \vec{r} - \vec{r'} |}
 \rho_s (\vec{r'})  \\
 \rho(\vec{r}) &=& e\sum_{j}Q_{j}\delta(\vec{r}-\vec{R}_{j})+e\sum_{k}\delta(\vec{r}-\vec{r}_{k})\\
 &-&e\sum_{l}\delta(\vec{r}-\vec{r}_{l}) \\
 \rho_{s}(\vec{r}) &=& \sum_{k}\delta(\vec{r}-\vec{r}_{k})+\sum_{l}\delta(\vec{r}-\vec{r}_{l})
\end{eqnarray*}
 One then  introduces a  second auxiliary field for the Yukawa component:
\begin{eqnarray*}
&&\exp \lbrace \frac {- \lambda_B C}{2} \int d\vec{r} \, d\vec{r'} 
\rho_s (\vec{r})
\frac {e^{-\kappa | \vec{r} - \vec{r'} |} }{| \vec{r} - \vec{r'} |} 
\rho_s (\vec{r'}) \rbrace    \\
&=&
\int D \psi (\vec{r}) \exp \lbrace
\frac {1}{8\pi\lambda_B} \int \psi (\Delta - \kappa^2) \psi d\vec{r} \\
&+&
i \sqrt{C}  \int \psi (\vec{r}) \rho_s (\vec{r}) \, d\vec{r} \rbrace
\end{eqnarray*}
Integrating over mobile charge positions as before yields an equivalent theory
in terms of two auxiliary fields $\chi,\psi$:
\begin{eqnarray*}
Z_{gc} &=&  \int {\cal D}\chi(\vec{r}) \int {\cal D}\psi(\vec{r})
\exp \lbrace \frac {\epsilon}{8 \pi \beta} \int \chi \Delta \chi d\vec{r} \\
&+& \frac {1}{8 \pi \lambda_B} \int \psi (\Delta - \kappa^2)
\psi d\vec{r} \\
&+& S_{\rm int}(\chi,\psi)+S_{\rm fix}(\vec{R}_{l}) \rbrace 
\end{eqnarray*}
with
\begin{eqnarray*}
S_{\rm int}(\chi,\psi) &=&  \frac {e^{\beta \mu_+}}{\lambda_+^3} \int d\vec{r}
e^{ie \chi (\vec{r}) + i \sqrt{C} \psi (\vec{r})
- V(\vec{r}) }  \\
&+&\frac {e^{\beta \mu_-}}{\lambda_-^3} \int d\vec{r}
e^{-ie \chi (\vec{r}) + i \sqrt{C} \psi (\vec{r})
- V(\vec{r}) } \\
S_{\rm fix}(\vec{R}_{l}) &=& ie \sum_l Q_l \chi (\vec{R_l})  
\end{eqnarray*}

 Similarly \cite{multipol}, one may derive an extension  of the formalism to deal with a multipolar gas, in 
which mobile charges also characterized by higher multipoles (dipole, quadrupole, etc):
\begin{eqnarray*}
  \rho(\vec{r})&\equiv& \sum_{j} e_j\delta(\vec{r}-\vec{r}_{j})  \\
  P^{a}(\vec{r}) &\equiv& \sum_{j} p_{j}^{a}\delta(\vec{r}-\vec{r}_{j}) \\
  Q^{ab}(\vec{r}) &\equiv& \sum_{j} q_{j}^{ab}\delta(\vec{r}-\vec{r}_{j}) 
\end{eqnarray*}
The total electrostatic energy for such a gas isgiven in terms of the effective charge density:
\begin{eqnarray*}
 \rho_{\rm eff}(\vec{r}) \equiv \rho(\vec{r})-\partial_{a}P^{a}(\vec{r})+\frac{1}{6}
\partial_{a}\partial_{b}Q^{ab}(\vec{r})+..
\end{eqnarray*}
\begin{eqnarray*}
 V_{\rm es}=\frac{1}{2}\int d^{3}\vec{r}d^{3}\vec{r}^{\prime}\rho_{\rm eff}(\vec{r})
\frac{1}{|\vec{r}-\vec{r}^{\prime}|}\rho_{\rm eff}(\vec{r}^{\prime})
\end{eqnarray*}
 As usual, we can localize the long-range interactions with an auxiliary field:
\begin{eqnarray*}
&&\hspace*{-0.35in}\exp{ \lbrace \frac {-\beta}{2} \int d^{3}\vec{r}d^{3}\vec{r'} 
\rho_{\rm eff} (\vec{r})
\frac {1}{| \vec{r} - \vec{r'} |} \rho_{\rm eff} (\vec{r'})}\rbrace    \\
   &=&\hspace*{-0.15in}
 \int D \chi (\vec{r}) \exp{  \lbrace     
\int(\frac {1}{8\pi\beta} \chi \Delta \chi + 
i\chi (\vec{r}) \rho_{\rm eff} (\vec{r})) d\vec{r}}\rbrace
\end{eqnarray*}

As an example, consider a gas of mobile permanent dipoles, $|\vec{p}_{j}|=\bar{p}$ ($Q^{ab}_{j},..=0$)
The configuration integral for a single dipole involves the average over orientations:
\begin{eqnarray*}
 Z_{\rm dip} &\equiv& \int \frac{d\hat{p}_{j}}{4\pi}e^{i\vec{p}_{j}\cdot\vec{\nabla}\chi(\vec{r}_{j})} \\
                     &=& \frac{\sin{(\bar{p}|\nabla\chi|)}}{\bar{p}|\nabla\chi|}
\end{eqnarray*}
 leading to {\em derivative interactions} in the effective field action:
\begin{eqnarray*}
S_{\rm int}(\chi) &=& e^{\beta\mu_{+}}\int e^{(ie\chi(\vec{r})-V(\vec{r}))}d\vec{r}\\
&+&
e^{\beta\mu_{-}}\int e^{(-ie\chi(\vec{r})-V(\vec{r}))}d\vec{r}  \\
&+&e^{\beta\mu_{d}}\int \frac{\sin{(\bar{p}|\nabla\chi(\vec{r})|)}}{\bar{p}|\nabla\chi(\vec{r})|}d\vec{r}
\end{eqnarray*}
so we are led to a  modified Poisson-Boltzmann equation, in which the mobile dipoles provide an effective
spatially varying dielectric, together with a correspondingly modified loop expansion.

\section{Weakly/Strongly Fluctuating Coulomb Gases}
 In a renormalizable local field theory like QCD, there is a natural dimensionless
coupling (typically, the running coupling at momentum scales relevant to the process
under consideration) which provides an expansion parameter for the saddle point 
expansion corresponding to covariant perturbation theory. In the case of
the effective field theories discussed above for Coulomb gas problems,
the validity of a perturbative loop expansion around the mean-field 
(Poisson-Boltzmann) theory depends on the ratio of two length scales:
\begin{enumerate}
\item The Bjerrum length $l_{B}\equiv$ distance between two mobile charges such
that pair electrostatic energy $=k_{B}T$.
\item The Gouy-Chapman length $\mu$, which depends in a more complicated way on the
geometry of the fixed charges relative to the mobile ones.
\end{enumerate}

\begin{figure}
\centerline{\psfig{figure=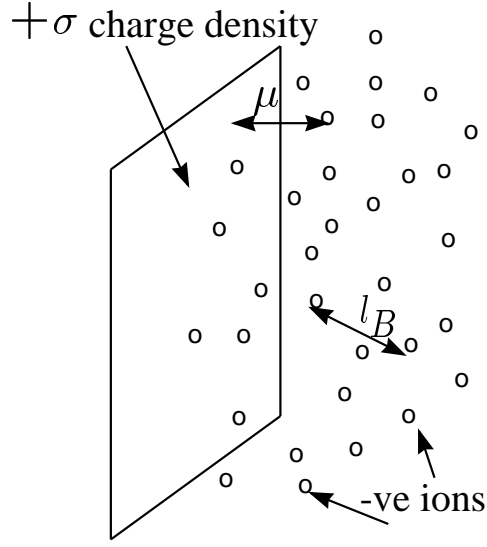,height=0.5\hsize}}
\caption{Coulomb gas near charged plate}
\end{figure}

As a  simple example (see \cite{MorNetz}), consider a gas of mobile ions of charge $-qe$ near a charged plate of charge density $+\sigma e$ (see Figure 1),
where the whole system is electrically neutral. For this problem, the Bjerrum and Gouy-Chapman
lengths are 
  \begin{eqnarray*}
  l_{B} = \frac{(qe)^{2}}{4\pi\epsilon k_{B}T},\;\;
  \mu = \frac{q}{2\pi l_{B}\sigma},\;\;
  \Xi \equiv \frac{l_{B}}{\mu}
  \end{eqnarray*}
In this case, the Gouy-Chapman length $\mu$  corresponds to the distance from the plate
at which an isolated mobile ion has electrostatic energy $k_{B}T$. The ratio $\Xi$ plays the role of the perturbative expansion parameter for this system, as can be seen
by rewriting the Hamiltonian for the system
\begin{eqnarray*}
 \frac{1}{k_{B}T}H=\sum_{j<k}\frac{l_{B}}{|r_{j}-r_{k}|}+\frac{2\pi l_{B}\sigma}{q}\sum_{j}z_{j}
 \end{eqnarray*}
in terms of distances rescaled to the Gouy-Chapman length, $r_{j}=\mu \bar{r}_{j}$:
 \begin{eqnarray*}
   \frac{1}{k_{B}T}H = \Xi\sum_{j<k}\frac{1}{|\bar{r}_{j}-\bar{r}_{k}|}+\sum_{j}\bar{z}_{j}
   \end{eqnarray*}
 Note  that the parameter $\Xi \propto q^{3}$ is extremely sensitive to the valence
   $q$- it often happens that we go from weak to strong fluctuations when q goes from 1 to 2
   (monovalent to divalent ions)!
   
    Unfortunately, in many interesting cases $\Xi >> 1$ and the perturbative
   loop expansion breaks down. Direct Monte Carlo simulation of the path integral
   Eq.(\ref{pathintegralzgc}) is not feasible:the action is
   complex and phase oscillations result in unmanageably large fluctuations (the infamous
   sign problem!).
  For such strongly coupled Coulomb gases (as for intrinsically nonperturbative field 
theories such as QCD in the infrared), we must resort to numerical simulation techniques,
but clearly one needs an alternative formulation where the effective action is real, allowing
the application of conventional Monte Carlo techniques.
  
  \section{Local Gauge Theory Approach to Coulomb Gases }

  Start from the Hamiltonian path integral for abelian gauge theory with external
  point charge sources: however $c\rightarrow\infty$ and magnetic effects and time-dependence
  are ignored (as well as spatial variation of the dielectric). Then the
  canonical partition function takes the form:

  \begin{eqnarray*}
  Z&=&\int\prod_{i=1}^{N}d\vec{r}_{i}{\cal D}\vec{E}(\vec{r})\prod_{\vec{r}}\delta(\vec{\nabla}
\cdot\vec{E}-\frac{1}{\epsilon}\rho(\vec{r})) \\
&&\cdot e^{-\frac{\beta\epsilon}{2}\int d\vec{r}\vec{E}^{2}
(\vec{r})} \\
 \rho(\vec{r}) &=& \sum_{i}e_{i}\delta(\vec{r}-\vec{r}_{i})+\sum_{l}Q_{l}\delta(\vec{r}-\vec{R}_{l})
\end{eqnarray*}
 The transverse, or curl part of the electric
field variable decouples from the charged particle dynamics via the
Helmholtz decomposition:
\begin{eqnarray*}
  \vec{E}&=& \vec{\nabla}\phi + \vec{\nabla}\times\vec{A}  \\ 
  \int d\vec{r}\,\vec{E}^{2} &=& \int d\vec{r}\, |\vec{\nabla}\phi|^{2}+\int d\vec{r}\,|\vec{\nabla}\times\vec{A}|^{2}
\end{eqnarray*}
The unphysical curl part of the electric field decouples from the gradient
part: only the latter sees the charge density $\rho$, so that the charges couple to 
yield the correct electrostatic
energy. In practice the functional integral is regulated on a spatial lattice, with mobile charged
particles associated with sites and electric field values with links (see Figure 2).

 Rescaling to lattice units, the Hamiltonian becomes:
\begin{eqnarray*}
    H = \frac{\hat{\beta}}{2}\sum_{l}\hat{E}_{l}^{2}
\end{eqnarray*}
and the Gauss' Law constraint takes the form:
\begin{eqnarray*}
       \sum_{l}\hat{E}_{l} = q_{i}
\end{eqnarray*}
for the sum of outgoing link fields  from any site containing a charged particle of
charge $q_{i}e$.

\begin{figure}
\centerline{\psfig{figure=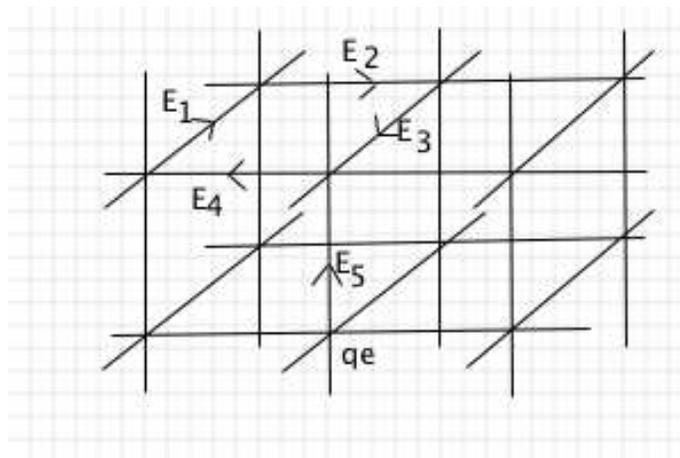,height=0.4\hsize}}
\caption{Electric Field Link Variables}
\end{figure}

 A simulation algorithm for this system is easily devised:
\begin{enumerate}
\item  Pick starting lattice locations (randomly) for the $N$ particles of charge $q_{i},i=1,..N$.
 Then solve  Gauss' Law for these fixed charge locations to obtain a starting
configuration of electric link field variables satisfying the Gauss constraint.  
\item Update the electric fields by shifting all link variables along a complete set of
independent closed paths by constant shifts.  In the simplest version of this, one simply
 considers all plaquettes (unit squares) on the lattice, shifting the 4 link fields ordered
around the plaquette by a random $\alpha, -\alpha_{\rm max}<\alpha<\alpha_{\rm max}$.
This preserves Gauss' law. In fact, for systems with constant dielectric, one
may use \cite{fftpaper} fast Fourier transform methods to effect a global update of the electric fields,
completely eliminating autocorrelations. 
\item Update particle locations by visiting in turn every site $ \vec{n}$ containing a charged particle
of charge $q_{i}$.
A particle move to the neighboring site $\vec{n}+\hat{\mu}$ in a random direction $\mu$ is then considered,
where the particle move is accompanied with a shift of the electric field $E_{l}$ on the link
$l=(\vec{n}\rightarrow\vec{n}+\hat{\mu})$
\begin{eqnarray*}
   E_{l} \rightarrow E_{l}-q_{i}
\end{eqnarray*}
to maintain Gauss' law. In practice \cite{CoalDunSedge1}, we need to couple particle moves to field changes on several
neighboring links to get reasonable acceptance rates.
\end{enumerate}
A classic example of a strongly coupled Coulomb gas is the case of two like charged plates
(see Figure 3),
with mobile counterions (rendering the system overall neutral) between the plates. For
appropriate choice of the charge density $\sigma$ on the plates, the system can be
switched from weakly to strongly coupled simply by doubling the valence of the counterions
(i.e. half as many doubly charged ions). For singly charged ions one finds, using
the local gauge simulation techniques described above, a positive
pressure, as shown in Figure 4 (also shown is the dependence on lattice
discretization for the same physical size system). Mean field theory, along the lines
discussed previously in Section 2, {\em always} gives a repulsive plate interaction of
this form. On the other hand, if we double the charge on each ion, leaving the plate
charge density fixed (and halving the number of counterions),
 one then finds an {\em attractive}
force between the like charged plates (negative pressure), implying a complete
breakdown of the saddle-point expansion of Section 2. The result of a numerical
simulation of this system using the local gauge method is shown in Fig. 5 (for details,
see \cite{CoalDunSedge1}), where the total pressure  (solid line) is seen to be
negative. These results agree with explicit (and painful) molecular dynamics simulations
\cite{molsymplate} of the like-charged-plate problem.

 \begin{figure}
\centerline{\psfig{figure=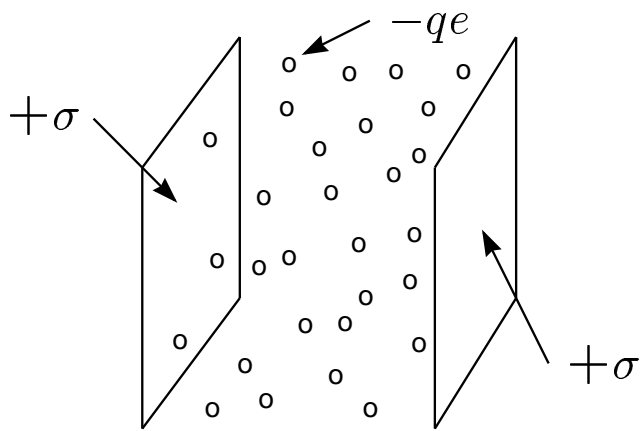,height=0.40\hsize}}
\caption{Coulomb gas between like charged plates}
\end{figure}

 \begin{figure}
\centerline{\psfig{figure=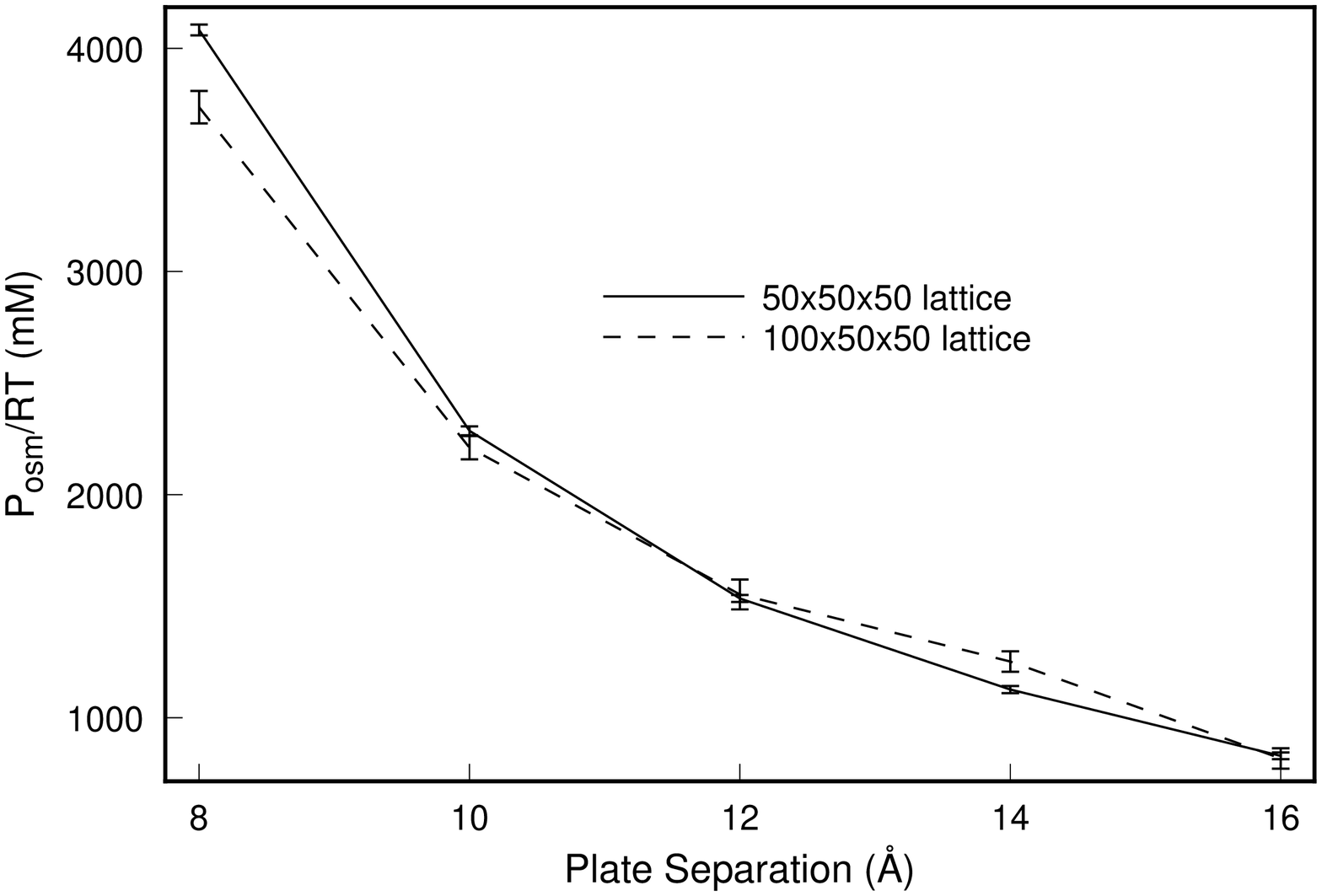,height=0.45\hsize}}
\caption{Plate pressure for univalent ions}
\end{figure}
 \begin{figure}
\centerline{\psfig{figure=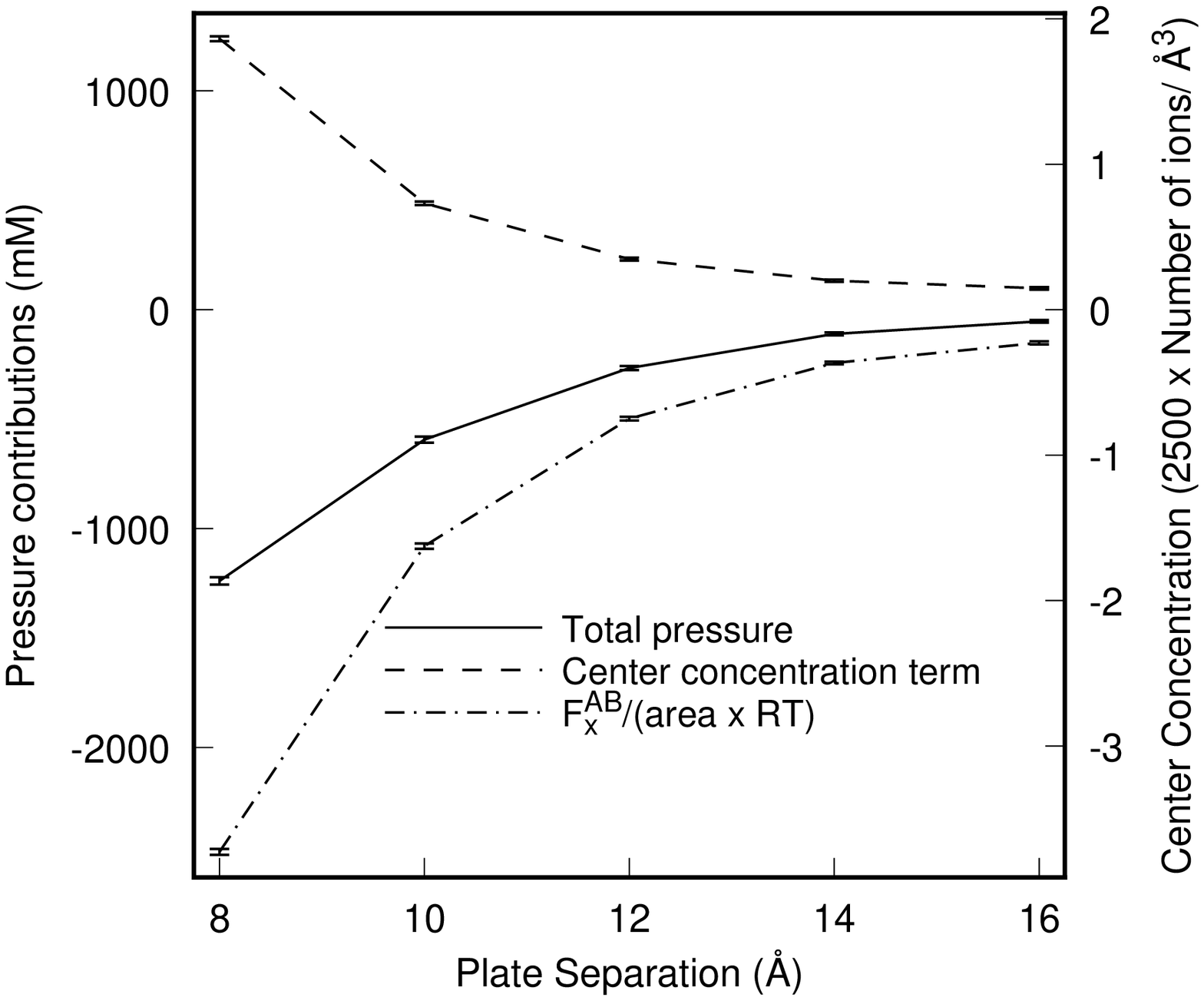,height=0.4\hsize}}
\caption{Plate pressure for divalent ions}
\end{figure}

\section{Inhomogeneous Dielectric Effects in Coulomb Gas Problems}
In realistic situations, the dielectric field  $\epsilon$ is NOT constant, but varies from 2-6 in the
interior of macromolecules to 80 in the surrounding medium (water)
\begin{eqnarray*}
  \label{eq:helmdecomp2}
   \vec{D}(\vec{r}) = -\epsilon(\vec{r})\vec{\nabla}\phi(\vec{r})+\vec{\nabla}\times\vec{A}(\vec{r})
                 \equiv \vec{D}^{||}(\vec{r})+\vec{D}^{\rm tr}(\vec{r})
\end{eqnarray*}
Maxwell's equations imply $\vec{D}^{\rm tr}=0$ but if this constraint is not 
explicitly imposed:
\begin{eqnarray*}
  \int d\vec{r}\frac{\vec{D}^{2}}{\epsilon(\vec{r})}=\int d\vec{r}\frac{\vec{D}^{||}(\vec{r})^{2}}{\epsilon(\vec{r})}
  +\int d\vec{r}\frac{\vec{D}^{\rm tr}(\vec{r})^{2}}{\epsilon(\vec{r})}
  \end{eqnarray*}
Ignoring the irrotational constraint, if we procede as previously:
  \begin{eqnarray*}
    Z^{\prime}&=&\int\prod_{i=1}^{N}d\vec{r}_{i}\prod_{\vec{r}}{\cal D}\vec{D}(\vec{r})\delta(\vec{\nabla}
\cdot\vec{D}-\rho(\vec{r}))\\
&&\cdot e^{-\frac{\beta}{2}\int d\vec{r}\vec{D}^{2}/\epsilon
(\vec{r})} \\
&& \hspace*{-0.45in}=\int\prod_{i=1}^{N}d\vec{r}_{i}\prod_{\vec{r}}{\cal D}\vec{D}^{||}(\vec{r}){\cal D}\vec{D}^{\rm tr}(\vec{r})
   \delta(\vec{\nabla}\cdot\vec{D^{||}}-\rho(\vec{r}))   \\
   &\cdot&e^{-\frac{\beta}{2}\int d\vec{r}\vec{D}^{||}(\vec{r})^{2}/\epsilon
(\vec{r})}e^{-\frac{\beta}{2}\int d\vec{r}\vec{D}^{\rm tr}(\vec{r})^{2}/\epsilon(\vec{r})} 
\end{eqnarray*}
The integral over $\vec{D}^{\rm tr}$ gives a spurious unphysical dependence on
$\epsilon(\vec{r})$- which can change dynamically in the course of the simulation ($\epsilon(\vec{r})$
depends implicitly on $\vec{r}_{i}$).
Regularizing the path integral with a spatial lattice, the exact form of the spurious $\epsilon$-dependence 
can be revealed simply by turning
off all free charges so the contribution from the longitudinal part vanishes:

\begin{eqnarray*}
 {\cal F}(\epsilon) &=& \int\prod_{n\mu}dD_{n\mu}\delta(\bar{\Delta}_{\mu}D_{n\mu})
 e^{-\frac{\beta}{2}\sum_{n\mu}D_{n\mu}^{2}/\epsilon_{n\mu} } \\
&&\hspace*{-0.65in}=\int\prod_{n}d\lambda_{n}\prod_{n\mu}dD_{n\mu}e^{i\sum_{n}\lambda_{n}\bar{\Delta}_{\mu}D_{n\mu}
  -\frac{\beta}{2}\sum_{n}D_{n\mu}^{2}/\epsilon_{n\mu}} \\
 &&\hspace*{-0.65in}=\int\prod_{n}d\lambda_{n}\prod_{n\mu}dD_{n\mu}e^{-i\sum_{n}D_{n\mu}\Delta_{\mu}\lambda_{n}
  -\frac{\beta}{2}\sum_{n}D_{n\mu}^{2}/\epsilon_{n\mu}} \\
 &&\hspace*{-0.45in}= C\prod_{n\mu}\sqrt{\epsilon_{n\mu}}\int \prod_{n}d\lambda_{n}e^{-\frac{1}{2\beta}\sum_{n\mu}\epsilon_{n\mu}
 (\Delta_{\mu}\lambda_{n})^{2}} \\
 &&\hspace*{-0.45in}= C^{\prime}\prod_{n\mu}\sqrt{\epsilon_{n\mu}}{\rm det}^{-\frac{1}{2}}(-\bar{\Delta}_{\mu}\epsilon_{\mu}\Delta_{\mu})
 \end{eqnarray*}
Evidently, in order to eliminate the spurious term, we need to include ${\cal F}^{-1}$ in the
path integral:
  \begin{eqnarray*}
   {\cal F}^{-1}(\epsilon) = \prod_{n\mu}e^{-\frac{1}{2}\sum_{n\mu}\log{(\epsilon_{n\mu}})}{\rm det}^{+\frac{1}{2}}(-\bar{\Delta}_{\mu}\epsilon_{\mu}\Delta_{\mu})
   \end{eqnarray*}
Including a {\em positive} power of the determinant of a local operator in
a path integral is precisely the problem we face in unquenched QCD!
\subsection{Eliminating transverse contributions with L\"uscher multiboson fields}
In the L\"uscher approach to unquenched QCD, one begins with a polynomial approximation to $1/s$ in the interval $[0,1]$;
a convenient choice is:
  \begin{eqnarray*}
 \frac{1}{s} &\simeq&   C\prod_{k=1}^{N}((s-\mu_{k})^{2}+\nu_{k}^{2})  \\
  \mu_{k} &=& \frac{1}{2}(1+\delta)(1-\cos{\frac{2\pi k}{2N+1}}) \\
  \nu_{k} &=& \sqrt{\delta}\sin{\frac{2\pi k}{2N+1}} 
  \end{eqnarray*}
implying  for the determinant (scaling the spectrum to $[0,1]$) a corresponding approximation:
 \begin{eqnarray*}
  {\rm det}^{+\frac{1}{2}}({\cal M}) \simeq \prod_{k=1}^{N} {\rm det}^{-\frac{1}{2}} (({\cal M}-\mu_{k})^{2}+\nu_{k}^{2})
  \end{eqnarray*}
This leads to  the corrected form for the path-integral:
   \begin{eqnarray*}
Z&=&\int d\vec{r}_{i}dD_{n\mu}d\phi^{(k)}_{n}\delta(\bar{\Delta}_{\mu}D_{n\mu}-\rho_{n})\\
  &&\cdot e^{-\frac{1}{2}\sum_{n\mu}\log{(\epsilon_{n\mu}})-\frac{\beta}{2}\sum_{n\mu}D_{n\mu}^{2}/\epsilon_{n\mu}}  \\
  &&\cdot e^{-\sum_{k=1}^{N}\phi^{(k)}(({\cal M}-\mu_{k})^{2}+\nu_{k}^{2})\phi^{(k)}}
 \end{eqnarray*}
The approximation only adequately describes the low eigenstates if one uses a sufficiently
large number $N$ of auxiliary fields, which limits its usefulness in QCD,
where the density of low eigenmodes of the Dirac operator is large, due to spontaneous
symmetry breaking.
Fortunately, in biophysical applications, the IR spectrum of the Poisson 
operator is sparse (unlike QCD) and we can get away with a small number of 
boson fields (as we shall se shortly, often $N$=4 is adequate).

\begin{figure}
\centerline{\psfig{figure=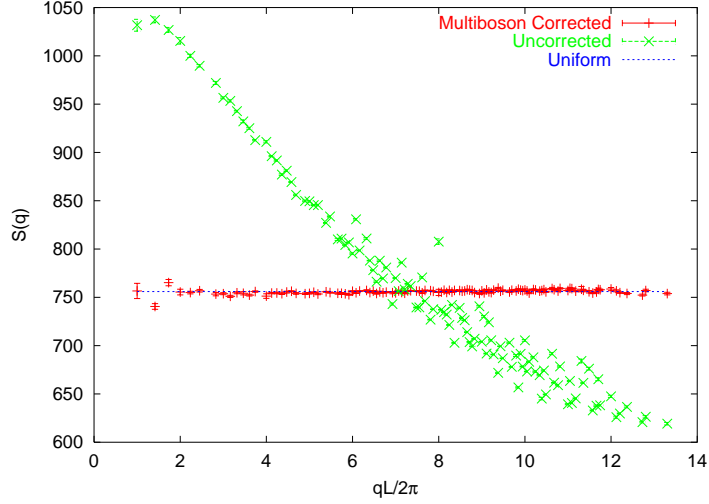,height=0.45\hsize}}
\caption{Structure Function for Neutral Dielectric Plasma}
\end{figure}

 \begin{figure}
\centerline{\psfig{figure=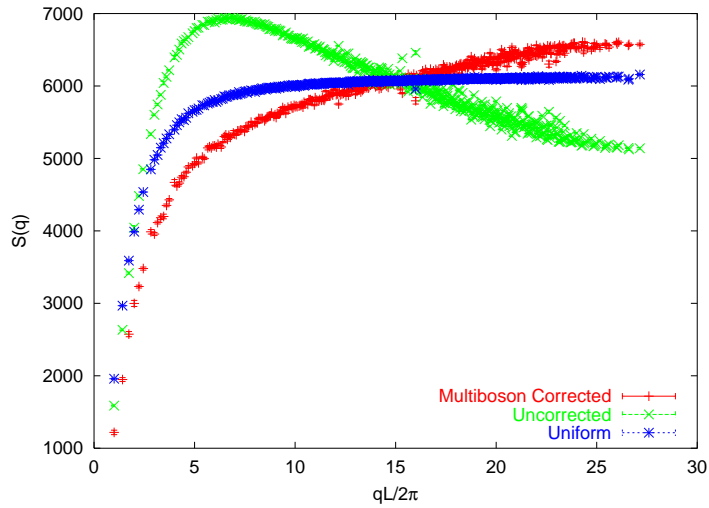,height=0.45\hsize}}
\caption{Structure Function for Charged Dielectric Plasma}
\end{figure}

  A useful testbed for the multiboson implementation of a Coulomb gas with dynamical dielectric
effects is the dielectric plasma, in which mobile particles (either neutral or charged) are
assigned a dielectric constant different from that of the ambient medium. It is convenient
to associate the dielectric field with links, each link being given a value depending on
whether a particle is present at either, neither or both of the end sites of the link
(for further details, see \cite{multibos1,multibos2}). An appropriate observable is the
structure factor $S(q)\equiv\int d\vec{r}e^{i\vec{q}\cdot\vec{r}}<\rho(\vec{r})\rho(0)>$:
for neutral particles, this must be $q$-independent. As a first test, consider a
 model with 1000 neutral dielectric particles,  on a 16$^3$ lattice, with a ratio of
particle to background dielectric given by
 $\epsilon_{\rm part}/\epsilon_{\rm  bg}$=0.2.
 
\begin{figure}
\centerline{\psfig{figure=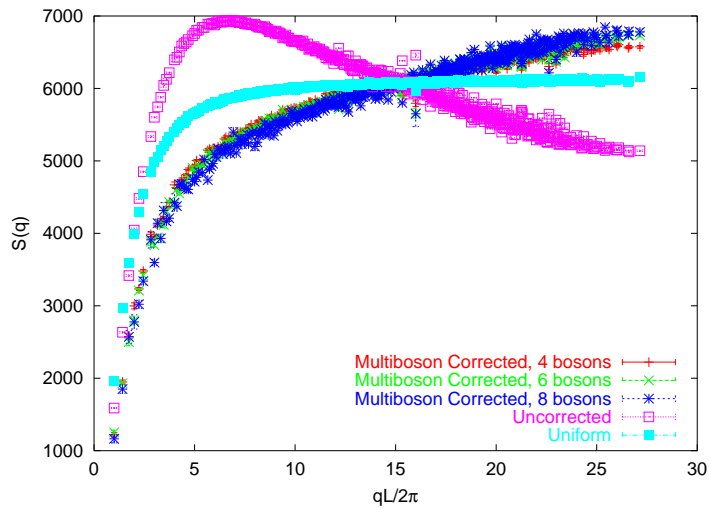,height=0.45\hsize}}
\caption{Dependence on number of multiboson fields}
\end{figure}

  As we see in Figure 6, once the multiboson fields are included, the measured structure factor
is indeed flat, to a very good approximation. A somewhat more realistic model is the
1 component charged dielectric plasma, on a  32$^3$ lattice, with  8000 particles, and
$\epsilon_{\rm part}/\epsilon_{\rm bg}$=0.05. In Figure 7 we show the results for such a
plasma, with and without the contribution from the multiboson fields. Evidently, the results
are qualitatively incorrect without taking into account the induced effects in the transverse
electric field. 
In Figure 8, we show the dependence on the number of multiboson fields used. As remarked
previously, it is remarkable how few fields are needed to achieve reasonable accuracy in this
model.

\section{Conclusions}

  The last 15 years has seen an extensive development in the theory of the
  thermodynamics of Coulomb gases,
  vastly extending the scope of treatable systems beyond the simplest cases to which
  elementary Debye-H\"uckel theory is applicable. The new methods treat a combined
  system of charged (and even multipolar) particles interacting with fields, and with
  both particles and fields realized on a discrete spatial lattice. Formally, the functional
  formalism for the grand-canonical partition function offers the greatest generality,
  but is restricted in usefulness to weakly coupled Coulomb gases. The local
  gauge formulation introduced by Maggs and Rossetto \cite{Maggs1}, and recently generalized
  to systems with dynamical dielectric by Coalson, Duncan and Sedgewick \cite{multibos1}
  allows the efficient simulation of strongly coupled Coulomb gases in a local formalism,
  circumventing the nonlocal Coulomb interaction which complicates, and frequently
  renders intractable,  the molecular dynamics
  approach.
  
  \section{Acknowledgements}
   It is a pleasure to acknowledge the numerous colleagues from whom I have learned
   an enormous amount  in the course of collaborations in the area of biophysics over
   the past 15 years: in particular, R. Coalson, R. Sedgewick, and S. Tsonchev, whose
   expertise and timely intervention have frequently averted scientific derailment. The research of 
   A. Duncan is supported in part by NSF contract PHY-0554660.

\end{document}